\documentstyle[preprint,tighten,aps,floats,epsfig]{revtex}
\def\slashed{{/}\mskip-10.0mu}
\def\pcircslash{\slashed {p\mskip -5mu ^{^\circ}}}
\begin{document}
\draft
\vskip 2cm

\title{The Critical Mass of Wilson Fermions:\\
A Comparison of Perturbative and Monte Carlo Results}

\author{E. Follana and H. Panagopoulos}
\address{Department of Physics, University of Cyprus, P.O. Box 20537,
Nicosia CY-1678, Cyprus \\
{\it email: }{\tt eduardo@dirac.ns.ucy.ac.cy, haris@ucy.ac.cy}}
\vskip 3mm

\date{\today}

\maketitle

\begin{abstract}

We calculate the critical value of the hopping
parameter, $\kappa_c$, in Lattice QCD with Wilson fermions, to two
loops in perturbation theory.

This quantity is an additive renormalization; as such, it is
characterized not only by the standard caveats regarding the
asymptotic nature of perturbative results, but also by a linear
divergence in the lattice spacing. Consequently, our calculation tests
rather stringently the limits of applicability of perturbation theory.

We compare our results to non perturbative evaluations of $\kappa_c$
coming from Monte Carlo simulations.

Finally, we apply a tadpole improvement technique on our results; this
shifts them quite favourably towards the non perturbative values.


\end{abstract}

\newpage


\section{Introduction}
\label{introduction}

In this paper we study the hopping parameter
in lattice QCD with Wilson fermions. In particular, we compute its
critical value to two loops in perturbation theory.

Wilson fermions are the most straightforward and widely used
implementation of fermionic actions on the lattice. This
implementation circumvents the fermion doubling problem by introducing a
higher derivative term with a vanishing classical continuum limit,
to lift unphysical propagator poles completely. 
At the same time, the action is strictly local, which is very advantageous
for numerical simulation.

The price one pays for strict locality and absence of doublers is, of
course, well known: The higher derivative term breaks chiral
invariance explicitly. Thus, merely setting the bare fermionic mass to
zero is not sufficient to ensure chiral symmetry in the quantum
continuum limit; quantum corrections introduce an additive
renormalization to the fermionic mass, which must then be fine tuned
to have a vanishing renormalized value. Consequently, the
hopping parameter $\kappa$, which is very simply related to the fermion mass,
must be appropriately shifted from its naive value, to recover chiral
invariance.

By dimensional power counting, the additive mass renormalization is
seen to be linearly divergent with the lattice spacing. This adverse 
feature of Wilson fermions poses an additional problem to a
perturbative treatment, aside from the usual issues related to lack
of Borel summability. Indeed, our calculation serves as a
check on the limits of applicability of perturbation theory, by
comparison with non perturbative results coming from Monte Carlo simulations.

Starting from our two-loop results, we also provide improved estimates
of the critical value of $\kappa$, by performing a resummation to all
orders of cactus diagrams \cite{cactus1}. These are tadpole-like
diagrams which are gauge 
invariant and dress the propagators and vertices in our
calculation. This improvement technique, among others, has so far been
applied mostly to the one-loop multiplicative renormalization of various
operators \cite{cactus2,cactus3}. It is interesting to explore to
what extent such methods 
lead to an improvement even in a sensitive case such as the one at hand.
We find that our improved estimates compare quite well with
Monte Carlo data also in this case.

The paper is organized as follows:
In Sec.~\ref{sec2} we define the quantities which we set out to
compute, and describe our calculation.
In Sec.~\ref{sec3} we present our results and compare with Monte Carlo
evaluations.
In Sec.~\ref{sec4} we obtain the improved estimates coming from cactus resummation.

\section{Formulation of the problem}
\label{sec2}

QCD with Wilson fermions on the lattice is described by the following
action (see, e.g., Ref. \cite{Rothe} for standard notation and conventions):

\begin{equation}
S_{\rm L} =  {1\over g_0^2} \sum_{x,\mu,\nu}
{\rm Tr}\left[ 1 - U_{\mu\nu}(x) \right]  +
 \sum_{i=1}^{N_f} \sum_{x,y} \bar{\psi}_i(x)D(x,y)\psi_i(y)
\label{latact}
\end{equation}
$U_{\mu\nu}(x)$ is the standard product of link variables $U_{x,y}$
around a plaquette in
the direction $\mu{-}\nu$, originating at point $x$, and $D(x,y)$ is given by:
\begin{equation}
D(x,y) = a m_B \, \delta_{x,y} + {1\over2}\,\sum_\mu \left[\gamma_\mu \, 
(U_{x,y}\delta_{x+\hat\mu,y} - U_{x,y}\delta_{x,y+\hat\mu}) - 
r\, (U_{x,y}\delta_{x+\hat\mu,y} -2 \delta_{x,y} +
U_{x,y}\delta_{x,y+\hat\mu})\right]
\end{equation}
As usual, $g_0$ denotes the bare coupling constant and $a$ is the
lattice spacing. The bare fermionic mass $m_B$ must be set to zero for
chiral invariance in the classical continuum limit.

The higher derivative term, multiplied by the Wilson coefficient $r$,
breaks chiral invariance. It vanishes in the classical continuum
limit; at the quantum level, it induces nonvanishing,
flavor-independent corrections to the fermion masses.

Numerical simulation algorithms usually employ the hopping parameter, 
\begin{equation}
\kappa\equiv{1\over 2\,m_B\,a + 8\,r}
\end{equation}
as a tunable quantity. Its critical value, at which chiral symmetry
is restored, is thus $1/8r$ classically, but gets shifted by quantum
effects.

The renormalized mass can be calculated in textbook fashion from the
fermion self--energy. Denoting by $\Sigma^L(p,m_B,g_0)$ the truncated,
one particle irreducible fermionic two-point function, we have for the
fermionic propagator:
\begin{eqnarray}
S(p)&=& {1\over i \,\pcircslash + m(p)} \sum_{k{=}0}^\infty
\left(\Sigma^L(p,m_B,g_0){1\over i \,\pcircslash + m(p)} \right)^k\nonumber\\
&=& \left[ i \,\pcircslash + m(p)- \Sigma^L(p,m_B,g_0)\right]^{-1}\\
{\rm where:}\qquad \pcircslash &=& \sum_\mu\gamma_\mu {1\over a} \sin(ap^\mu), \quad m(p) =
m_B + {2r\over a} \sum_\mu \sin^2(ap^\mu/2).\nonumber
\end{eqnarray}

Requiring that the renormalized mass vanish, leads to:
\begin{equation}
S^{-1}(0) = 0 \qquad\Longrightarrow\qquad m_B = \Sigma^L(0,m_B,g_0)
\end{equation}
The above is a recursive equation for $m_B$, which can be solved order
by order in perturbation theory.

\bigskip
We write the loop expansion of $\Sigma^L$ as:

\begin{equation}
\Sigma^L(0,m_B,g_0) = g_0^2 \, \Sigma^{(1)} + g_0^4 \, \Sigma^{(2)} + \cdots
\end{equation}
Fig. I shows the two diagrams contributing to the 1-loop
result $\Sigma^{(1)}$. The fermion mass involved in these diagrams
must be set to its tree level value, $m_B\to 0$. The $i^{\rm th}$
diagram gives a contribution of the form
$\frac{N^2 - 1}{N} \, c^{(1)}_i$, where $c^{(1)}_1,c^{(1)}_2$ are
numerical constants. 

\bigskip
\hskip4.0cm\psfig{figure=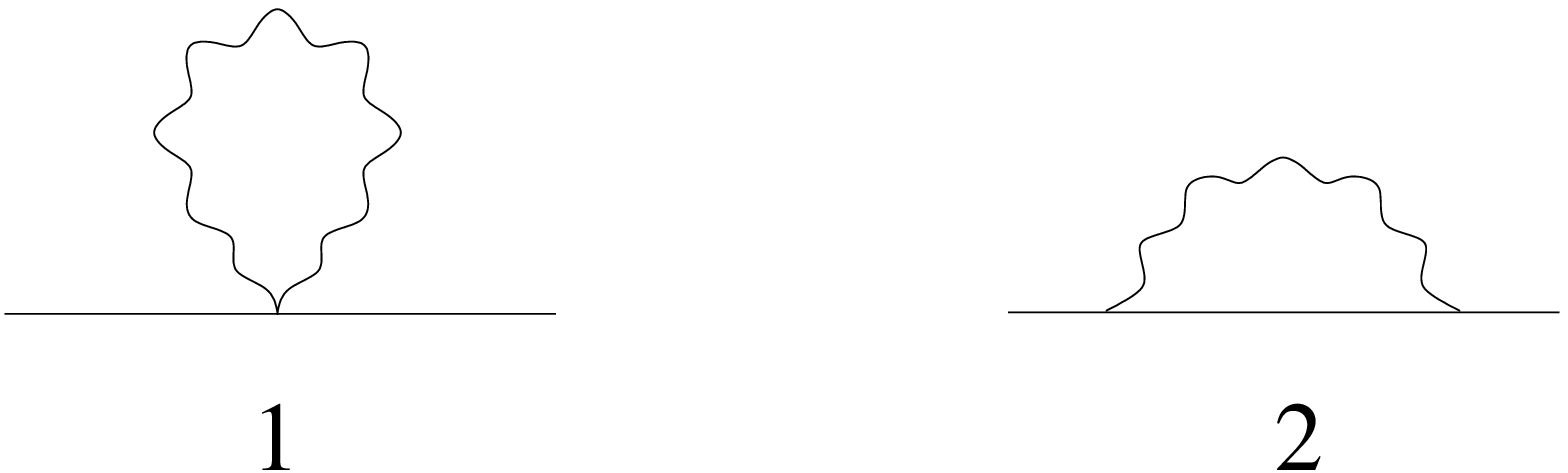,height=2truecm}\hskip1.0cm
\nopagebreak\medskip

\noindent
{\small FIGURE I.\ \ One-loop diagrams contributing to $\Sigma^L$.
Wavy (solid) lines represent gluons (fermions).}

\bigskip
A total of 26 diagrams contribute to the 2-loop quantity $\Sigma^{(2)}$, shown in
 Fig. II. Genuine 2-loop diagrams must again be evaluated at
 $m_B\to 0$; in addition, one must include to this order the 1-loop diagram
 containing an ${\cal O}(g_0^2)$ mass counterterm (diagram 23).
The contribution of
each diagram can be written in the form

\begin{equation}
(N^2 - 1) (c^{(2)}_{1,i} + \frac{c^{(2)}_{2,i}}{N^2} + 
\frac{N_f}{N} c^{(2)}_{3,i})
\label{c4}
\end{equation}
\noindent
where $c^{(2)}_{1,i},c^{(2)}_{2,i},c^{(2)}_{3,i}$ are numerical
constants. Certain sets of diagrams, corresponding to renormalization of loop
 propagators, must be evaluated together in
order to obtain an infrared-convergent result: these are diagrams 7+8+9+10+11,
12+13, 14+15+16+17+18, 19+20, 21+22+23.

\bigskip
\hskip2.0cm\psfig{figure=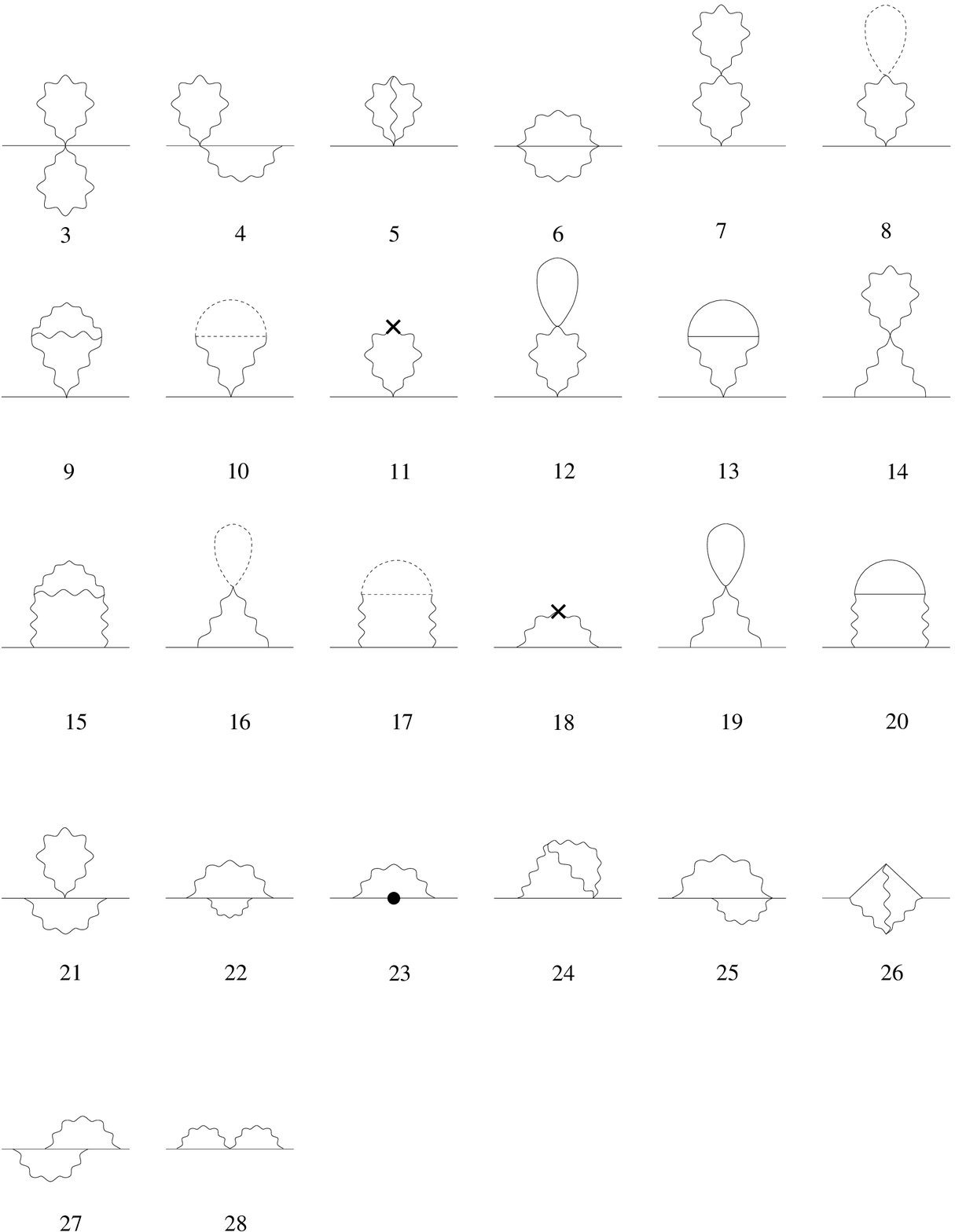,height =14truecm}\hskip1.0cm
\medskip

\noindent
{\small FIGURE II.\ \ Two-loop diagrams contributing to $\Sigma^L$.
Wavy (solid, dotted) lines represent gluons (fermions,
ghosts). Crosses denote vertices stemming from the measure part of the
action; a solid circle is a fermion mass counterterm.}

\section{Numerical Results}
\label{sec3}

The evaluation of the diagrams in this computation requires very extensive analytical
work. To this end, we use a Mathematica package which we
have developed for symbolic manipulations in lattice perturbation
theory (see, e.g., Ref. \cite{C-F-P-V-98}). Applied to the present
case, this package allows us to perform in a rather straightforward way the
following tasks: Contraction among the appropriate
vertices; simplification of color/Dirac matrices; use of trigonometry
and momentum symmetries for reduction to a more compact, canonical
form; automatic generation of highly optimized
Fortran code for the loop integration of each type of expression.

The integrals, typically consisting
of a sum over a few hundred trigonometric products, are then performed
numerically on lattices of varying finite size $L$.
Our programs perform extrapolations of each expression to a broad spectrum
of functional forms of the type: $\sum_{i,j} e_{ij} (\ln L)^j/L^i$,
analyze the quality of each extrapolation using a variety of criteria
and assign statistical weights to them, and finally produce a quite
reliable estimate of the systematic error. Taking $L\le 28$ leads to a
sufficient number of significant digits in the results we present.

One important consistency check can be performed on those diagrams
which are separately IR divergent; taken together in groups, as listed
below Eq.~(\ref{c4}), they give a finite and very stable extrapolation.

We present below the numerical values of the constants $c^{(1)}_i,
c^{(2)}_{1,i}, c^{(2)}_{2,i}, c^{(2)}_{3,i}$. These constants depend
only on the Wilson parameter $r$; following common practice, we set
$r=1$.

Table \ref{tab1} contains the contributions to the 1-loop quantity
$\Sigma^{(1)}$. The total 1-loop result  is

\begin{equation}
\Sigma^{(1)} = \frac{N^2 - 1}{N} \, (-0.162857058711(2))
\end{equation}

This result is known in the literature (see, e.g.,
Ref. \cite{Montvay}, p. 246, and references contained therein).

The contributions to the 2-loop quantity $\Sigma^{(2)}$ are presented
in table \ref{tab2}. The total 2-loop result is

\begin{equation}
\Sigma^{(2)} = (N^2 - 1) ( -0.017537(3) + \frac{1}{N^2} \, 0.016567(2) +
\frac{N_f}{N} \, 0.00118618(8) )
\label{sigma2}
\end{equation}

In order to make a comparison with numerical simulations, let us set
$N=3$, $N_f=2$ in the above; we obtain

\begin{equation}
\Sigma^{(1)}(N{=}3, N_f{=}2) = -0.434285489897(5)
\end{equation}
\begin{equation}
\Sigma^{(2)}(N{=}3, N_f{=}2) = -0.11925(3)
\end{equation}

In Table \ref{tab3} we compare the final results for $m_c^{(1)} =
g_0^2 \Sigma^{(1)}$ and $m_c^{(2)} = g_0^2 \Sigma^{(1)} + g_0^4
\Sigma^{(2)}$ with numerical 
simulation data at values of $\beta = 6/g_0^2$ equal to $5.6$
\cite{SESAM} and $5.5$ \cite{SCRI}. Also included in this table are the
improved results obtained with the method described in the following
Section. For easier reference, Table \ref{tab4} presents our results
in terms of the critical hopping parameter $\kappa_c = 1/(2\, m_c \, a +
8\, r)$.

\section{Improved Perturbation Theory}
\label{sec4}

In order to obtain improved estimates from lattice perturbation
theory, one may perform a resummation to all orders of the so-called
``cactus'' diagrams~\cite{cactus1,cactus2,cactus3}. 
Briefly stated, these
are gauge--invariant tadpole diagrams which become disconnected if any one of their
vertices is removed. The original motivation of this procedure is the well known
observation of ``tadpole dominance'' in lattice perturbation theory.
In the following we refer to Ref.~\cite{cactus1} for 
definitions and analytical results.

Since the contribution of standard tadpole diagrams is not gauge invariant,
the class of gauge invariant diagrams we are considering needs further specification.
By the Baker-Campbell-Hausdorff (BCH) formula, the product of link variables
along the perimeter of a plaquette can be written as
\begin{eqnarray}
U_{x,\mu\nu}&& = 
e^{i g_0 A_{x,\mu}} e^{i g_0 A_{x+\mu,\nu}} e^{-i
g_0 A_{x+\nu,\mu}} e^{-i g_0 A_{x,\nu}} \nonumber \\
&&=\exp\left\{i g_0 (A_{x,\mu} + A_{x+\mu,\nu} -
A_{x+\nu,\mu} - A_{x,\nu}) + {\cal O}(g_0^2) \right\} \nonumber \\
&&=\exp\left\{ i g_0 F_{x,\mu\nu}^{(1)} +  i g_0^2
F_{x,\mu\nu}^{(2)} + {\cal O}(g_0^4)
\right\}
\end{eqnarray}
The diagrams that we propose to resum to all orders are the cactus
diagrams made of vertices containing $F_{x,\mu\nu}^{(1)}\,$.
Terms of this type come from the pure gluon  part of
the lattice action.  These diagrams dress the transverse gluon
propagator $P_A$ leading to an improved propagator $P_A^{(I)}$,
which is a multiple of the bare transverse one:
\begin{equation}
P_A^{(I)} = {P_A\over 1-w(g_0)},
\label{propdr}
\end{equation}
where the factor $w(g_0)$ will depend on $g_0$ and
$N$, but not on the momentum.
The function $w(g_0)$ can be extracted by an
appropriate algebraic equation that
 has been derived in Ref.~\cite{cactus1} and that can be easily
solved numerically; for $SU(3)$, $w(g_0)$ satisfies:
\begin{equation}
u \, e^{-u/3} \, \left[u^2 /3 - 4u +8\right]  = 2 g_0^2, \qquad 
u(g_0) \equiv {g_0^2 \over 4 (1-w(g_0))}.
\end{equation}
The vertices coming from the gluon part of the action, Eq.~(\ref{latact}),
get also dressed using a procedure similar to the one leading to 
Eq.~(\ref{propdr}) \cite{cactus1}.
Vertices coming from the fermionic action stay unchanged, since their 
definition contains no plaquettes on which to apply the linear BCH formula.

One can apply  the resummation of cactus diagrams to the calculation of
additive and multiplicative renormalizations of lattice operators.
Applied to a number of cases of interest~\cite{cactus1,cactus2}, this procedure yields 
remarkable improvements when compared with the available nonperturbative estimates.
As regards numerical comparison with other improvement schemes, such as
boosted perturbation theory~\cite{Parisi-81,L-M-93}, cactus
resummation fares equally well on all the cases
studied~\cite{cactus3}. 

One advantageous feature of cactus resummation, in comparison to other
schemes of improved perturbation theory, is the possibility of
systematically incorporating higher loop diagrams. The present
calculation best exemplifies this feature, as we will now show.

Dressing the 1-loop result is quite straightforward: the fermionic
propagator and vertices stay unchanged, and only the gluon propagator
gets simply multiplied by $1/(1-w(g_0)$. The resulting values:
$m^{(1)}_{c,\,\rm dressed}$ and $\kappa^{(1)}_{c,\,\rm dressed}$, are
shown in Tables \ref{tab3} and \ref{tab4}, respectively. It is worth
noting that these values already fare better than the much more
cumbersome undressed 2-loop results.

We now turn to dressing the 2-loop results. Here, one must take care
to avoid double counting: A part of diagrams 7 and 14 has already been
included in dressing the 1-loop result, and must be explicitly
subtracted from $\Sigma^{(2)}$ before dressing. Fortunately, this part
(we shall denote it by $\Sigma^{(2)}_{\rm sub}$)
is easy to identify, as it necessarily includes all of the $1/N^2$ part
in $\Sigma^{(2)}$. A simple exercise in contraction of $SU(N)$
generators shows that $\Sigma^{(2)}_{\rm sub}$ is proportional to
$(2N^2-3)(N^2-1)/(3N^2)$. There follows immediately that:
\begin{equation}
\Sigma^{(2)}_{\rm sub} = -0.016567 (2N^2-3)(N^2-1)/(3N^2)
\end{equation}
(cf. Eq. \ref{sigma2}).

A further complication is presented by gluon vertices. While the
3-gluon vertex dresses by a mere factor of $(1-w(g_0))$, the dressed
4-gluon vertex contains a term which is not simply a multiple of its
bare counterpart (see Appendix C of Ref.~\cite{cactus1}). 
Once again, however, we are fortunate: this term
must be dropped, being precisely
the one which has already been taken into account in dressing the
1-loop result, while the
remainder dresses in the same way as the 3-gluon vertex. In
conclusion, cactus resummation applied to the 2-loop quantity
$\Sigma^{(2)}$ leads to the following rather simple recipe:
\begin{equation}
m^{(2)}_{c,\,\rm dressed} = \Sigma^{(1)}\, {g_0^2\over 1-w(g_0)} + 
(\Sigma^{(2)}- \Sigma^{(2)}_{\rm sub})\, {g_0^4\over [1-w(g_0)]^2}
\end{equation}

\medskip
The end results, $m^{(2)}_{c,\,\rm dressed}$ and
$\kappa^{(2)}_{c,\,\rm dressed}$, are included in Tables \ref{tab3}
and \ref{tab4}. Comparing with the Monte Carlo estimates, we see a
definite improvement over non-dressed values. At the same time, a
sizeable discrepancy still remains, as was expected from start. This
discrepancy sets a benchmark for lattice perturbation theory;
multiplicative renormalizations, calculated to the same order and
improved by cactus dressing, are expected to be much closer to their
exact values. We hope to return to these calculations in a future
publication.


\newpage
\begin{table}[ht]
\begin{center}
\begin{minipage}{6cm}
\caption{Coefficients $c^{(1)}_i$. $r=1$.
\label{tab1}}
\begin{tabular}{cr@{}l}
\multicolumn{1}{c}{$i$}&
\multicolumn{2}{c}{$c^{(1)}_i$} \\
\tableline \hline
1 &-0&.15493339023106 \\
2 &-0&.007923668480(2) \\
\end{tabular}
\end{minipage}
\end{center}
\nobreak
\begin{center}
\begin{minipage}{15cm}
\caption{Coefficients $c^{(2)}_{1,i}$, $c^{(2)}_{2,i}$,
$c^{(2)}_{3,i}$. $r=1$.
\label{tab2}}
\begin{tabular}{cr@{}lr@{}lr@{}l}
\multicolumn{1}{c}{$i$}&
\multicolumn{2}{c}{$c^{(2)}_{1,i}$} &
\multicolumn{2}{c}{$c^{(2)}_{2,i}$} &
\multicolumn{2}{c}{$c^{(2)}_{3,i}$} \\
\tableline \hline
3    &  0&.002000362950707492 & -0&.00030005444260612375&  0 &\\
4    &  0&.00040921361(1)     & -0&.00061382041(2)      &  0 &\\
5    &  0  &                  &  0 &                    &  0 &\\
6    & -0&.0000488891(8)      &  0&.000097778(2)        &  0 &\\
7+8+9+10+11  & -0&.013927(3) &  0&.014525(2)           &  0 &\\
12+13 &  0 &                   &  0 &                    &  0&.00079263(8) \\
14+15+16+17+18 & -0&.005753(1)&  0&.0058323(7)          &  0 &\\
19+20&  0 &                   &  0&                     &  0&.000393556(7) \\
21+22+23    & 0&.000096768(4) &  -0&.000096768(4)       &  0 &\\
24    &  0 &                   &  0&                     &  0 &\\
25   &  0&.00007762(1)        &  -0&.00015524(3)        &  0 &\\
26   & -0&.00040000(5)        &  0 &                    &  0 &\\
27   &  0  &                  &  -0&.000006522(1)       &  0 &\\ 
28   &  0&.0000078482(5)      &  -0&.000015696(1)       &  0 &\\
\end{tabular}
\end{minipage}
\end{center}
\nobreak
\begin{center}
\begin{minipage}{11cm}
\caption{$m_c^{(1)}$ and $m_c^{(2)}$. $N=3$, $N_f = 2$, $r=1$.
\label{tab3}}
\begin{tabular}{lr@{}lr@{}l}
\multicolumn{1}{c}{}&
\multicolumn{2}{c}{$\beta = 5.5$}&
\multicolumn{2}{c}{$\beta = 5.6$}\\
\tableline \hline
$m_c^{(1)}$ & -0&.473765988978(5) & -0&.465305882032(5) \\ 
$m_c^{(2)}$ & -0&.61568(3) & -0&.60219(3) \\
$m_{c,\,\rm dressed}^{(1)}$ & -0&.658392964276(7) & -0&.640695803036(7) \\
$m_{c,\,\rm dressed}^{(2)}$ & -0&.76323(6) &   -0&.73997(6) \\
Simulation &  -0&.8975 & -0&.8446  \\
\end{tabular}
\end{minipage}
\end{center}
\nobreak
\begin{center}
\begin{minipage}{11cm}
\caption{$\kappa_c^{(1)}$ and $\kappa_c^{(2)}$. $N=3$, $N_f = 2$, $r=1$.
\label{tab4}}
\begin{tabular}{lr@{}lr@{}l}
\multicolumn{1}{c}{}&
\multicolumn{2}{c}{$\beta = 5.5$}&
\multicolumn{2}{c}{$\beta = 5.6$}\\
\tableline \hline
$\kappa_c^{(1)}$ & 0&.1417943331149(4) & 0&.1414549557367(4) \\ 
$\kappa_c^{(2)}$ & 0&.147740(3) & 0&.147154(3) \\
$\kappa_{c,\,\rm dressed}^{(1)}$ & 0&.1496286052353(6) & 0&.1488403462990(6) \\
$\kappa_{c,\,\rm dressed}^{(2)}$ & 0&.154475(6) & 0&.153373(6) \\
Simulation & 0&.16116(15)  & 0&$.158507^{+41}_{-44}$  \\
\end{tabular}
\end{minipage}
\end{center}
\end{table}
\end{document}